\documentclass[twocolumn,showpacs,preprintnumbers,amsmath,amssymb]{revtex4}

\usepackage{bm}

\begin{document}

\title{General Relativity Histories Theory}

\author{Ntina Savvidou}
\email{k.savvidou@imperial.ac.uk}

\affiliation{Physics Department, Theory Group, Imperial College,
South Kensington, London SW72BZ, UK \\
and\\
Physics Department, University of Patras, 26500 Patras, Greece}

\date{\today}

\begin{abstract}
The canonical description is based on the prior choice of a
spacelike foliation, hence making a  reference to a spacetime
metric. However, the metric is expected to be a dynamical,
fluctuating quantity in quantum gravity. After presenting the
developments in the History Projection Operator histories theory in
the last seven years---giving special emphasis on the novel temporal
structure of the formalism---we show how this problem can be solved
in the histories formulation of general relativity. We implement the
$3+1$ decomposition using metric-dependent foliations which remain
spacelike with respect to all possible Lorentzian metrics. This
allows us to find an explicit relation of covariant and canonical
quantities which preserves the spacetime character of the canonical
description. In this new construction we have a coexistence of the
spacetime diffeomorphisms group $ \rm{Diff}(M)$ \emph{and} the Dirac
algebra of constraints.
\end{abstract}


 \maketitle


\section{Introduction}

In this work we present the application of ideas of consistent
histories to general relativity, and its potential implications for
the quantisation of the theory--regarding in particular the emphasis
on spacetime concepts of the histories. We show how the temporal
structure of the Histories Projection Operator formalism led to some
very important consequences regarding the physical results of
canonical general relativity.

One of the major approaches to the quantisation of gravity is the
canonical one, either in its original form---involving
geometrodynamic variables---or in terms of the loop variables,
introduced via the connection formulation of canonical general
relativity.

The canonical quantisation involves: \vspace{0.2cm}

i) the identification of a Hilbert space on which the canonical
commutation relations---or some other appropriate algebraic
structure---can be implemented, thereby defining the kinematical
variables of quantum gravity. The Hilbert space is chosen to allow
the representation of the constraints of the Hamiltonian description
in terms of self-adjoint operators, preserving the classical Dirac
algebra of constraints.

ii) then, one has to find the zero eigenspace of the constraint
operators, in order to define the physical Hilbert space. This is
the scope of the original Dirac quantisation of constrained systems:
variations are usually employed in the case of gravity (or special
models), because the constraint operators are not expected to have a
discrete spectrum.

However the canonical quantisation scheme suffers also from serious
problems, both on technical level and conceptual level. For example,
we encounter problems in constructing the Hilbert space, writing the
constraint operators, and finding their spectrum. Also, the fact
that general relativity is a generally covariant theory raises grave
doubts about the conceptual adequacy of the canonical method of
quantisation.

Furthermore, the equations of general relativity are covariant with
respect to the action of the diffeomorphisms group ${\rm Diff}(M)$,
of the spacetime manifold $M$. This does not pose great difficulties
in the classical theory, since once the equations of motion are
solved the Lorentzian metric on $M$ can be used to implement
concepts like causality and spacelike separation. In quantum theory
however, such notions as causality and spacelike separation are
lost, because the geometry of spacetime is expected to be subject to
quantum fluctuations.

This creates problems even at the first step of the quantisation
procedure, namely the definition of the canonical commutation
relations. The canonical commutation relations are defined on a
`spacelike' surface, however, a surface is spacelike with respect to
some particular spacetime metric $g$, which is itself a quantum
observable that is expected to fluctuate.

The prior definability of the canonical commutation relations is not
merely a mathematical requirement: in a generic quantum field theory
the canonical commutation relations implement the principle of {\em
microcausality}: namely that field observables that are defined in
spacelike separated regions commute. However, if the notion of
spacelikeness is also dynamical, it is not clear in what way this
relation will persist.

A spacelike foliation is necessary for the implementation of the
$3+1$ decomposition and the definition of the Hamiltonian. Again we
are faced with the question of how to: reconcile  the requirement of
spacelikeness  with the expectation that different metrics will take
part in the quantum description.

Even more, one may question whether the predictions of the resulting
quantum theories are independent of the choice of foliation. The
Hilbert space of the quantum theory, which it is constructed
canonically, is not straightforwardly compatible with the ${\rm
Diff}(M)$ symmetry. In the canonical theory, the symmetry group is
the one generated by the Dirac algebra of constraints, which is
mathematically distinct from the ${\rm Diff}(M)$ group.

In effect, different choices of foliation lead {\em a priori} to
different quantum theories and there is no guarantee that these
quantum theories are unitarily equivalent (or physically equivalent
in some other generalised sense).

The canonical description cannot provide an  answer to these
questions, because once the foliation is employed for the $3+1$
decomposition, its effect is lost, and there is no way of relating
the predictions corresponding to different foliations.

The above are serious problems, which challenge the validity of the
canonical approach towards the description of a {\em generally
covariant} theory of quantum gravity.

Finally, the problem which is perhaps most well known, is the
problem of time. The Hamiltonian of general relativity is a
combination of the first class constraints, hence it vanishes on the
reduced state space. It is expected also to vanish on the physical
Hilbert space constructed in the quantisation scheme. This means
that there is no notion of time evolution in the space of true
degrees of freedom. More than that, the notion of \emph{time as
causal ordering} seems to be lost. In contrast, the tensorial
expressions of the equations of motion are ${\rm Diff}(M)$-invariant
in the Lagrangian formalism. It seems very natural, therefore, to
wish for a theory that combines the virtues of both formalisms: the
Lagrangian, and the Hamiltonian. Such a theory is the history
projection operator-histories (HPO) scheme, which offers the
possibility of handling the ideas of space and time in significantly
new ways.

\bigskip

The consistent histories scheme was developed by Griffiths and
Omn\'es\cite{CoHis1} as an interpretation of quantum theory for
closed systems. Gell-Mann and Hartle\cite{CoHis2} elaborated this
scheme in the case of quantum cosmology.

Further development came from Isham and Linden and collaborators in
the HPO histories scheme\cite{Ish94,IL94,IL95,ILSS98}, in which they
were able to represent histories by projection operators on a
suitable Hilbert space, thus emphasising the temporal quantum logic
of the frame work.

An important feature of the HPO histories is the augmented temporal
structure\cite{S99}, which allows us to mathematically implement the
distinction between time as a parameter of kinematics and as a
parameter of dynamics. It is of great significance that, in the
context of classical canonical general relativity this distinction
provides a framework in which the spacetime diffeomorphism group
\emph{coexists} with the Dirac algebra of constraints. This is a
very significant result: it implies that there is a central role for
{\em spacetime\/} concepts, as opposed to the domination by spatial
ideas in the normal canonical approaches to quantum gravity. More
important, it allows a kinematical description in which different
choices of the direction of time coexist, in a way that always
preserves the spacetime character of the theory.

The (general relativity) constraints, depend on the foliation
functional. This leads to the natural question, whether physical
results depend upon this choice.  The solution of the constraints
determines a reduced phase space for histories, which has an
explicit dependence on the foliation. In\cite{Sav03a, Sav03b} it was
showed that the action of the spacetime diffeomorphism group
intertwines between different reduction procedures. Moreover, if one
requires that a specific physical `equivariance condition' is
satisfied by the foliation functional, then {\em the reduced state
space is invariant under the action of the diffeomorphism group}.
This is a completely novel result, which has been made possible only
by the incorporation of general relativity into the histories
formalism.

\bigskip

The histories approach to general relativity suggests a new,
spacetime focussed, approach to quantum gravity that is
characterised by two features that are not implemented in any of the
existing, direct, quantum gravity schemes.

First, the Lorentzian metric is quantised, as a direct analogue of
the way the `external' quantum field arises in the history approach
to scalar quantum field theory\cite{Sav01,IS01}. On the other hand,
in the conventional canonical quantum gravity schemes only the {\em
spatial\/} metric on a three-surface is quantised.

Second, the history scheme incorporates intrinsically the basic
symmetry of general relativity, namely \textsl{general covariance},
as manifested by the existence of a realisation of the group of
spacetime diffeomorphisms, and under whose action the history
analogue of the canonical Hamiltonian constraints are invariant.


\section{Consistent Histories Preliminary}

The consistent histories formalism was originally developed by
Griffiths and Omn\'es\cite{CoHis1}, as an interpretation of quantum
theory for closed systems.

Gell-Mann and Hartle\cite{CoHis2} elaborated this scheme in the case
of quantum cosmology--the Universe being regarded as a closed
system. They emphasised in particular that a theory of quantum
gravity that is expected to preserve the spacetime character of
general relativity would need a quantum formalism in which the
irreducible elements are temporally extended objects, namely
histories.

The basic object in the consistent histories approach is a history
\begin{equation}
\alpha:= ( {\hat{\alpha}}_{t_{1}} , {\hat{\alpha}}_{t_{2}} ,... ,
{\hat{\alpha}}_{t_{n}}),
\end{equation}
which is a time-ordered sequence of properties of the physical
system, each one represented by a single-time projection operator on
the standard Hilbert space. We notice that the emphasis is given on
histories rather than states at a single time.

The probabilities and the dynamics are contained in the decoherence
functional, a complex-valued function on the space of histories
\begin{equation}
d_{H , \rho } ( \alpha\,, \beta) = tr ( {{\tilde{C}}_
{\alpha}}^{\dagger} {\rho}_{t_0} C_{\beta} ),
\end{equation}
where $\rho_{t_0}$ is the initial quantum state and where

\begin{equation}
{\tilde{C}}_{\alpha}:= U (t_0, t_1 )\hat{\alpha}_{t_1} U (t_1, t_2 )
... U (t_{n-1}, t_n) \hat{\alpha}_{t_n} U (t_n, t_0 )
\end{equation}
is the class operator that represents the history $\alpha$.

When a set of histories satisfies a decoherence condition,
\begin{equation}
d_{\mathcal{H} , \rho } ( \alpha\,, \beta) = 0 \hspace{0.5cm}  then
\hspace{0.3cm} \alpha \,, \beta \hspace{0.1cm} in\; the\;
consistent\; set,
\end{equation}
which means that we have zero interference between different
histories, then it is possible to consistently assign probabilities
to each history in that set; it is called a \emph{consistent set}.

Then we can assign probabilities to each history in the consistent
set
\begin{equation}
d_{\mathcal{H} , \rho } ( \alpha\,, \alpha) = Prob( \alpha ;
\rho_{t_0} ) = tr ( {{\tilde{C}}_ {\alpha}}^{\dagger} {\rho}_{t_0}
C_{\alpha} ).
\end{equation}

\section{History Projection Operator-Basic elements}

In the History Projection Operator(HPO) approach to consistent
histories theory the emphasis is given on the \emph{temporal}
quantum logic.

A history is represented by a tensor product of projection operators
\begin{equation}
\hat{\alpha}:=\hat{\alpha}_{t_{ 1}} \otimes \hat{\alpha}_{t_{2}}
\otimes ... \otimes \hat{\alpha}_{t_{n}},
\end{equation}
each operator $\hat{\alpha}_{t_{i}}$ being defined on a copy of the
single-time Hilbert space ${\cal{H}}_{t_i}$ at that time $t_i$ and
corresponding to some property of the system at the same time
indicated by the t-label. Therefore a history is \textsl{itself} a
genuine projection operator defined on the history Hilbert space
${\cal V}_n$, which is a tensor product of the single-time Hilbert
spaces
\begin{equation}
{\cal{V}}_{n}:= {\cal{H}}_{t_{1}}\otimes {\cal{H}}_{t_{2}}\otimes
... \otimes {\cal{H}}_{t_{n}}.
\end{equation}

In order to define continuous time histories, we do not take the
continuous limit of the tensor product of Hilbert spaces, as it
cannot be properly defined. The \textsl{history group}, which is a
generalised analogue of the canonical group of standard quantum
theory for elementary systems, was employed\cite{IL95} in order to
construct the continuous-time history Hilbert space.

For example, for a particle moving on a line the single-time
canonical commutation relations
\begin{eqnarray}
 [\,\hat{x}, \hat{x}^{\prime}\,]&=&0= [\,\hat{p},\hat{p}^{\prime}\,]
\\
 {[}\,\hat{x}, \hat{p}\,]&=& i \hbar
\end{eqnarray}
become the history group that it is described by the following
history commutation relations, defined at unequal moments of time
\begin{eqnarray}
[\, \hat{x}_{t} , \hat{x}_{t^{\prime}}\,] &=& 0 = {[}\, \hat{p}_{t}
, \hat{p}_{t^{\prime}}\,]  \label{xtxt'} \\
{[}\, \hat{x}_{t} , \hat{p}_{t^{\prime}}\,] &=& i \hbar
\delta(t-t^{\prime}). \label{xtpt'}
\end{eqnarray}
The key idea in the definition of the history group is, that, the
spectral projectors of the generators of its Lie algebra represent
propositions about phase space observables of the system.

The notion of a `{\em continuous\/} tensor product'---and hence
`{\em continuous\/} temporal logic'---arises via a representation of
the history algebra. In order to describe discrete-time histories we
have to replace the delta function, on the right-hand side of Eq.
(\ref{xtpt'}), with the Kronecker delta.

Propositions about histories of the system are associated with
projectors on history Hilbert space. We must clarify here that the
operator $x_t$ refers to the position of the particle at a specific
fixed moment of time $t$. As we shall see in the following section,
the novel temporal structure that was later introduced\cite{S99},
allowed the interpretation of the index $t$ as the index that does
not refer to dynamics---it is \emph{not} the parameter of time
evolution---it is the label of the temporal quantum logic, in the
sense that it refers to the time a proposition about momentum or
position is asserted.

It is important to remark that physical quantities are naturally
time-averaged in this scheme. The smeared form of the history
algebra
\begin{eqnarray}
 [\, \hat{x}_{f} , \hat{x}_{g}\,] &=& 0 = {[}\, \hat{p}_{f} ,
 \hat{p}_{g}\,] \\
 {[}\, \hat{x}_{f} , \hat{p}_{g}\,] &=& i \hbar (f \,,g), \label{W3}
\end{eqnarray}
where: $(f\,,g)= \int_{-\infty}^{\infty} dt\; f(t) g(t) $, resembles
that of an one-dimensional quantum field theory and therefore
techniques from quantum field theory may be used in the study of
these representations. Analogous versions of the history group have
been studied for other field theories\cite{Sav01, Burch03}.

The existence of a properly defined Hamiltonian operator $H$ is
proved to uniquely select the physically appropriate representation
of the history algebra, therefore the definition of the
time-averaged energy operator $H$ is crucial for the formalism.

\section{HPO-temporal structure}
In order to study the temporal structure of the HPO theory we use
the model of a one-dimensional simple harmonic oscillator, however
the results are generalised appropriately for other systems.

One of the crucial problems for the development of the HPO theory
was the lack of a clear notion of time evolution, in the sense that,
there was no natural way to express the time translations from one
time slot---that refers to one copy of the Hilbert space ${\cal
H}_t$---to another one, that refers to another copy ${\cal
H}_{t^{\prime}}$. The introduction of the history group allowed the
definition of continuous-time histories and led to `time-averaged'
physical observables, however any notion of dynamics was lost and
the theory was brought to a hold.

The situation changed after the introduction of a new idea
concerning the notion of time: the distinction between dynamics and
kinematics corresponds to the mathematical distinction between the
notion of `time evolution' from that of `time ordering' or `temporal
logic time'. The distinction proved very fruitful for the
development of the history theory, leading in particular to the
results of general relativity.

The crucial step in the identification of the temporal structure of
the theory was the definition in \cite{Sav01} of the action operator
$S$---a quantum analogue of the Hamilton-Jacobi functional
\cite{Dirac}, written as
\begin{equation}
 S_{\kappa}:= \int^{+\infty}_{-\infty}\!\!dt\;( p_{t}\dot{x}_{t}- \kappa(t)H_{t}) ,
 \label{Def:op_S}
\end{equation}
where $\kappa(t)$ is an appropriate test function.

The first term of the action operator $S_{\kappa}$  Eq.\
(\ref{Def:op_S}) is identical to the kinematical part of the
classical phase space action functional. This `Liouville' operator
is formally written as
\begin{equation}
 V:= \int^{\infty}_{-\infty}\! dt\,(p_{t}\dot{x_{t}})  \label{liou}
\end{equation}
so that
\begin{equation}
 S_{\kappa} = V - H_{\kappa}.
\end{equation}

The `average-energy' operator
\begin{eqnarray*}
\hat{H}_\kappa=\int_{-\infty}^{\infty}dt\, \kappa(t) \hat{H}_t; \ \
H_t&:=&{p_{t^2}\over 2m}+{m\omega^2\over 2}x_t^2
\end{eqnarray*}
is also smeared in time by smearing functions $\kappa(t)$. The
Hamiltonian operator may be employed (here for the special case
$\kappa(t)=1$) to define {\em Heisenberg\/} picture operators for
the smeared operators like $x_f$
\begin{eqnarray*}
    \hat{x}_{f}(s):=e^{\frac{i}{\hbar} s \hat{H}}\; \hat{x}_{f}\;
    e^{-\frac{i}{\hbar} s \hat{H}}  \nonumber
\end{eqnarray*}
where $f=f(t)$ is a smearing function.

Hence $\hat{H}_\kappa$ generates transformations with respect to the
Heisenberg picture parameter $s$, therefore, \textsl{$s$ is the time
label as it appears in the implementation of dynamical laws}
\begin{eqnarray*}
 e^{\frac{i}{\hbar} \tau \hat{H}}\: \hat{x}_{ f}(s)\:
 e^{-\frac{i}
 {\hbar} \tau \hat{H}} &=& \hat{x}_{ f}(s\!+\!\tau).
\end{eqnarray*}

The novel feature in this construction is the definition of the
`Liouville' operator $\hat{V}$, which is the quantum analogue of the
kinematical term in the classical phase space action functional. The
Liouville operator generates transformations with respect to the
time label $t$---as it appears in the history algebra, hence,
\textsl{$t$ the label of temporal logic or the label of kinematics}
\begin{eqnarray*}
 \hspace*{-0.1cm}e^{\frac{i}{\hbar} \tau \hat{V}}\: \hat{x}_{ f}(s)
  e^{-\frac{i}
 {\hbar} \tau \hat{V}} = \hat{x}_{f^\prime}(s)\,, \hspace{1cm} f^\prime(t)=
 f(t\!+\!\tau).
 \end{eqnarray*}

We must emphasise here the distinction between the notion of time
evolution from that of logical time-ordering. The latter refers to
the {\em temporal ordering} of logical propositions in the
consistent histories formalism. The corresponding parameter $t$ does
not coincide with the notion of physical time,---as it is measured
for instance by a clock. It is an abstraction, which keeps from
physical time only its {\em ordering} properties, namely that it
designates the sequence at which different events happen--the same
property that is kept by the notion of time-ordered product in
quantum field theory. Making this distinction about time, it is
natural to assume that in the HPO histories one may not use the same
label for the time evolution of physical systems and the
time-ordering of events. The former concept incorporates also the
notion of a clock, namely it includes a {\em measure} of time
duration, as something distinct from temporal ordering.

The realisation of this idea on the notion of time was possible in
this particular framework because of the logical structure of the
theory, as it was originally introduced in the consistent histories
formalism and as it was later recovered as temporal logic in the HPO
scheme.

One may say then that the definition of these two operators, $V$ and
$H$, implementing time translations, signifies the
\emph{distinction} between the kinematics and the dynamics of the
theory.

However a crucial result of the theory is that
\textsl{$\hat{S}_{\kappa}$ is the physical generator of the time
translations in histories theory}, as we can see from the way it
appears in the decoherence functional and hence the physical
predictions of the theory.

\section{Classical histories}

The HPO scheme and especially the history group suggests a
reformulation of classical mechanics in the language of histories,
which will prove very fruitful in the case of general relativity.

We consider the space of classical histories $\Pi  = \{ \gamma
\mid\gamma  : \mathbb{R} \rightarrow \Gamma \}$ as paths on the
single-time classical phase space $\Gamma$. We equip the history
space with a symplectic structure $t\rightarrow (x_t , p_t )$
corresponding to the following Poisson brackets
\begin{eqnarray*}
  \{x_t , x_{t^{\prime}} \} &=& 0  \\
  \{p_t , p_{t^{\prime}} \} &=& 0  \\
  \{x_t , p_{t^{\prime}} \} &=& \delta (t , t^{\prime})
\end{eqnarray*}
 where
\begin{eqnarray*}
 x_t : \Pi & \rightarrow & \mathbb{R}  \\
 \gamma & \mapsto & x_t (\gamma) := x( \gamma (t)).
\end{eqnarray*}

The classical Hamilton equations may be written in terms of the
Liouville function $V$ and the smeared Hamiltonian function $H$,
which are the classical analogues of the corresponding operators we
defined for the quantum theory
\begin{eqnarray*}
 \{ F_t , V \}_{\Pi} (\gamma_{cl}) = \{ F_t , H \}_{\Pi}
 (\gamma_{cl}),
\end{eqnarray*}
 where
\begin{eqnarray*}
 V(\gamma) := \int dt p_t \dot{x}_t\,, \hspace{1cm} \{F_t,
 V\}=\dot{F}_t.
\end{eqnarray*}
It follows the important conclusion that the solutions to the
classical equations of motion are the specific paths that remain
invariant under the symplectic transformations generated by the
action $S$ for all functions $F_t$
\begin{eqnarray*}
  \{\, F \,, S\, \}_{\Pi} ({\gamma}_{cl}) = 0, \label{lap}
\end{eqnarray*}
where $ S = V-H $. The Eq.\ (\ref{lap}) is essentially the histories
analogue of the least action principle.


\section{General relativity histories formalism}

Next we study the HPO formalism in the case of general
relativity\cite{Sav03a, Sav03b}. We show that the novel temporal
structure of HPO---that distinguishes between the kinematics and the
dynamics of a theory---suggests a spacetime description that is
immediately related to the canonical one.

Let us consider a 4-manifold $M$ with topology $\Sigma \times R$,
for a three-manifold $\Sigma$. We define the covariant history space
\begin{equation}
\Pi^{cov} = T^{*}{\rm LRiem}(M)
\end{equation}
as the cotangent bundle of the space of all Lorentzian, globally
hyperbolic four metrics on M and where ${\rm LRiem}(M)$ is the space
of all Lorentzian four-metrics.

$\Pi^{cov}$ is equipped with a symplectic structure with symplectic
form $\Omega$
\begin{eqnarray}
 \Omega = \int \!d^4X \, \delta \pi^{\mu\nu}(X) \wedge \delta
 g_{\mu\nu}(X)  \nonumber
\end{eqnarray}
where $X \in M$, $g_{\mu\nu} \in L {\rm Riem}(M)$ and $\pi^{\mu\nu}$
its `conjugate' variable and where $\delta g_{\mu\nu}$ is a one-form
on $\Pi^{cov}$ and $\delta$ represents the exterior derivative.

Or else with the covariant Poisson brackets algebra, on $\Pi^{cov}$
\begin{eqnarray} \{g_{\mu\nu}(X)\,,
\,g_{\alpha\beta}(X^{\prime})\}&=& 0 =\{\pi^{\mu\nu}(X)\,,
\,\pi^{\alpha\beta}(X^{\prime})\}
\nonumber \\
\{g_{\mu\nu}(X)\,, \,\pi^{\alpha\beta}(X^{\prime})\} &=&
\delta_{(\mu\nu )}^{\alpha\beta} \,\delta^4 (X, X^{\prime}) ,
\label{covgpi} \nonumber
\end{eqnarray}
where ${{\delta}_{(\mu\nu)}}^{\alpha\beta} :=
\frac{1}{2}({\delta_\mu}^\alpha {\delta_\nu}^\beta +
{\delta_\mu}^\beta {\delta_\nu}^\alpha )$.

The physical meaning of $\pi$ can be understood after the $3+1$
decomposition of $M$ in which it will be related to the canonical
conjugate momenta.

\subsection{The representation of the group $\rm Diff(M)$.}

The relation between the group of spacetime diffeomorphisms
$Diff(M)$ and the Dirac algebra of constraints has been an important
matter of discussion in quantum gravity. We have showed that in this
formalism of general relativity there exists a representation of the
group of spacetime diffeomorphisms \textsl{together} with the Dirac
algebra of constraints.

$\Pi^{cov}$ carries a symplectic action of the
 ${\rm Diff}(M)$  group, with generator defined for any vector field
 $W$ on $M$
\begin{eqnarray} V_W:=\int \!d^4X
\,\pi^{\mu\nu}(X)\,{\cal L}_W g_{\mu\nu}(X) \label{V_W}
\end{eqnarray}
where ${\cal L}_W$ denotes the Lie derivative with respect to $W$.

The functions $V_W$ satisfy the Lie algebra of ${\rm Diff}(M)$
\begin{eqnarray}  \{\, V_{W_1}\,,
V_{W_2}\,\} = V_{ [ W_1 , W_2 ]} \nonumber
\end{eqnarray}
{{where $[ W_1 , W_2 ]$ is the Lie bracket between vector fields
$W_1$ and $W_2$ on the manifold $M$.}}


\subsection{Relation between spacetime and canonical description}

Next we study the relation between the covariant description and the
standard canonical one.

We must emphasise here that the spacetime description we presented
is kinematical---in the sense that we do not start from a Lagrangian
formalism and from this deduce the canonical constraints. We rather
start from the histories canonical general relativity and we show
that this formalism is augmented by a spacetime description that
carries a representation of the spacetime $\rm Diff(M)$ group.

In the standard canonical formalism we introduce a spacelike
foliation ${\cal E}:\mathbb{R}\times\Sigma\rightarrow M$ on $M$,
with respect to a fixed Lorentzian four-metric $g$. Then the
spacelike character of the foliation function implies that the
pull-back of the four metric on a surface $\Sigma$ is a Riemannian
metric with signature $+++$. In the histories theory we obtain a
path of such Riemannian metrics $t\mapsto h_{ij}(t,\underline{x})$
each one defined on a copy of $\Sigma_t$ with the same $t$ label.

However a foliation cannot be spacelike with respect to \textsl{all}
metrics $g$ and in general, for an arbitrary metric $g$ the pullback
of a metric ${\cal{E}}^* g$ \textsl{is not} a Riemannian metric on
$\Sigma$.

This point reflects a major conceptual problem of quantum gravity:
the notion of `spacelike' has \textsl{no a priori} meaning in a
theory in which the metric is a non-deterministic dynamical
variable; in absence of deterministic dynamics, the relation between
canonical and covariant variables appears rather puzzling. In
classical general relativity this is not a problem because
`spacelikeness' refers to the metric that solves the equations of
motion. In quantum gravity however where one expects metric
fluctuations the notion of spacelikeness is problematic.

In histories theory this problem is addressed by introducing the
notion of a \textsl{metric dependent} foliation ${\cal{E}}[g]$,
defined as a map ${\cal{E}}[g]: {\textsl{LRiem}(M)}\mapsto
{\textsl{Fol}{M}} $, that assigns to each Lorentzian metric a
foliation that is \textsl{always} spacelike with respect to that
metric. Then we use the metric dependent foliation ${\cal{E}}[g]$ to
define the canonical decomposition of the metric $g$ with respect to
the canonical three-metric $h_{ij}$, the lapse function $N$ and the
shift vector $N^i$ as
\begin{eqnarray}
 h_{ij}(t,\underline{x})\!\!\! &:=&\!\!\!
 {\cal{E}}^{\mu}_{,i}(t,\underline{x};g]\,
 {\cal{E}}^{\nu}_{,j}(t,\underline{x};g]\,
 g_{\mu\nu}({\cal{E}}(t,\underline{x};g])\nonumber \\
 N_{i}(t,\underline{x})\!\!\! &:=&\!\!\!
 {\cal{E}}^{\mu}_{,i}(t,\underline{x};g]\,
 {\dot{\cal{E}}}^{\nu}(t,\underline{x};g]\, g_{\mu\nu}({\cal{E}}
 (t,\underline{x};g])\nonumber \\ \hspace{-3.5cm}
 -\!\!N^{2}(t,\underline{x})\!\!\! &:=&\!\!\!
 {\dot{\cal{E}}}^{\mu}(t,\underline{x};g]\,
 {\dot{\cal{E}}}^{\nu}(t,\underline{x};g]\,
 g_{\mu\nu}({\cal{E}}(t,\underline{x};g])\!\!\!- \!\!N_{i} N^{i}
\nonumber
\end{eqnarray}
Defined in this way $h_{ij}$ is \emph{always} a Riemannian metric,
with the correct signature.

In the histories theory therefore, \emph{the $3+1$ decomposition
preserves the spacetime character of the canonical variables\/}, a
feature that we may expect to hold in a theory of quantum gravity.


\subsection{Relation between $\Pi^{cov}$ and $\Pi^{can}$}
With the introduction of the metric-dependent foliation we can then
write the symplectic form $\Omega$, on the space of canonical
general relativity histories description $\Pi^{can}$, (using an
equivalent canonical form of $\Omega$),
\begin{eqnarray}
 \Omega &=& \int \!d^4X \, \delta \pi^{\mu\nu} \wedge \delta
 g_{\mu\nu}  \nonumber\\
 &=& \int\!d^3xdt (\delta {\pi}^{ij}\wedge\delta h_{ij} +
 \delta {p} \wedge\delta N
 + \delta {{p}}_{i}\wedge \delta N^{i}),
 \nonumber
\end{eqnarray}
by introducing conjugate momenta for the three-metric ${\pi}^{ij}$,
the lapse function $p$ and the shift vector ${{p}}_{i}$.

Thus we prove the equivalence of the covariant history space
$\Pi^{cov}= (T^{*}{\rm LRiem}(M)$ with the space of paths on the
canonical phase space of general relativity
\begin{equation}
\Pi^{can}\!\! = \!{\times}_{t} (T^{*}{\rm Riem}({\Sigma}_{t}) \times
T^{*}Vec({\Sigma}_{t}) \times T^{*}C^{\infty}({\Sigma}_{t})),
\end{equation}
where ${\rm Riem}({\Sigma}_{t})$ is the space of all Riemannian
three-metrics on the surface ${\Sigma}_{t}$, $Vec({\Sigma}_{t})$ is
the space of all vector fields on ${\Sigma}_{t}$, and
$C^{\infty}({\Sigma}_{t})$ is the space of all smooth scalar
functions on ${\Sigma}_{t}$.

\subsection{Canonical description}
The canonical history space of general relativity $\Pi^{can}$ is the
Cartesian product of the cotangent bundles of the space of
Riemannian three-metrics $\textsl{Riem}{{\Sigma}_t}$, the space of
vector fields on $\Sigma$ and the space of all scalar functions on
$\Sigma$. Hence, $\Pi^{can}$ is a suitable subset of the Cartesian
product of  copies of the phase space $\Gamma$ of standard canonical
general relativity\\
\begin{equation}
\Pi^{can}\subset {\times}_t {\Gamma}_t , \hspace{1cm}\Gamma_t =
\Gamma(\Sigma_t).
\end{equation}

A history therefore is any smooth map
\begin{equation}
t\mapsto\!\! (h_{ij}(t,\underline{x}),\pi^{kl}(t,\underline{x}),
N^{i}(t,\underline{x}), p_{i}(t,\underline{x}), N(t,\underline{x}),
p(t,\underline{x})).
\end{equation}

We obtain the history version of the canonical Poisson brackets from
the covariant Poisson brackets
\begin{eqnarray} \{ h_{ij}
(t,\underline{x})\,, \pi^{kl} (t^{\prime} , \underline{x}^{\prime}
)\} &=& {{\delta}_{(ij)}}^{kl}\, \delta ( t , t^{\prime})\, \delta^3
( \underline{x} ,
\underline{x}^{\prime}) \label{GR3b} \nonumber\\
\{N(t,\underline{x}), p(t',\underline{x}')\} &=& \delta(t,t')
\delta^3(\underline{x}', \underline{x'}) \label{cansNp}\nonumber\\
\{ N^i(t, \underline{x}), p_j(t',\underline{x}') \} &=& \delta^i_j
 \delta(t,t')\delta^3(\underline{x}', \underline{x'})
\label{canvNp} \nonumber\\
\{h_{ij} (t,\underline{x})\,, h_{kl} ( t^{\prime} ,
\underline{x}^{\prime} ) \} &=&0= \{ \pi^{ij} (t,\underline{x})\,,
\pi^{kl} ( t^{\prime} ,\underline{x}^{\prime} ) \} \nonumber\\
\{N(t,\underline{x}), N(t', \underline{x'}) \} &=&0=
\{p(t,\underline{x}), p(t', \underline{x'}) \}\nonumber \\
\{N^i(t,\underline{x}), N^j(t', \underline{x'}) \} &=&0=
\{p_i(t,\underline{x}), p_j(t', \underline{x'}) \} \nonumber
\end{eqnarray}
where we have defined ${{\delta}_{(ij)}}^{kl} :=
\frac{1}{2}({\delta_i}^k {\delta_j}^l + {\delta_i}^l {\delta_j}^k
)$. All quantities$N,N^i,p$ and $p_i$ have vanishing Poisson
brackets with $\pi^{ij}$ and $h_{ij}$.


\subsection{Invariance transformations of covariant and canonical descriptions}
The generators of the diffeomorphism group $\rm Diff(M)$, defined as
Eq.\ (\ref{V_W}), act on the spacetime or covariant variables in a
natural way, generating spacetime diffeomorphisms
\begin{eqnarray}
 \{\, g_{\mu\nu}(X)\,, V_W \,\} &=& {\cal L}_W g_{\mu\nu}(X)
 \nonumber\\[2pt]
 \{\, \pi^{\mu\nu}(X)\,, V_W \,\} &=& {\cal L}_W \pi^{\mu\nu}(X).
 \nonumber
\end{eqnarray}

The coexistence of the spacetime and the canonical variables allows
one to write the history analogue of the canonical constraints. The
canonical description leads naturally to a one-parameter family of
super-hamiltonians $t\mapsto{\cal{H}}_{\bot} ( t, \underline{x})$
and super-momenta $t\mapsto {\cal{H}}_i ( t, \underline{x} )$,
\begin{eqnarray} \hspace{-1cm}{\cal H}_\perp(t,\underline
x)\!\!\!&:=&\!\!\!\kappa\!^2
h\!^{-1/2}(t,\underline{x})(\pi^{ij}(t,\underline{x})\pi_{ij}
(t,\underline{x})\!\! \nonumber\\
&-& \!\!\frac{1}{2} (\pi_i{}^i)\!^2(t,\underline{x})) -
\kappa\!^{-2}h^{1/2}(t,\underline{x})
R(t,\underline{x})  \nonumber\\
\hspace{-1cm} {\cal{H}}^i(t,\underline x)\!\!\!&:=&\!\!\! - 2
{\nabla}_{\!\!j} \pi^{ij}(t,\underline x),\nonumber
\end{eqnarray}
where $\kappa^2 = \frac{8 \pi G}{c^2}$ and the nabla $\nabla$
denotes the covariant derivative. We prove that they satisfy a
history version of the Dirac algebra.

We may also write the constraints in a covariant form:
\begin{eqnarray} {\cal{H}}[ \vec{L}
]\!\!\!\! &=& \!\!\!\int\!\! d^4X \!(\bar{E} \pi)^{\mu \nu} {\cal
L}\!{_L} g_{\mu \nu}\! + \!2\!\! \int\!\! d^4X \!(\bar{E} \pi)^{\mu
\nu} n_{\mu} n^{\rho}
{\cal L}\!_{L} g_{\rho \nu} \nonumber\\
{\cal H}\!_{\perp}[L]\!\!\!\! &=&\!\!\!\! \int\!\!\! d^4\! X
\!\left[\kappa^2\! \frac{N}{\sqrt{-g}} \frac{1}{2}G_{\mu \nu \rho
\sigma} (\bar{E}\pi)^{\mu \nu}\! (\bar{E}\pi)^{\rho \sigma}
\!\!\!-\! \!\!\kappa\!\!^{-2} \frac{\sqrt{-g}}{N}
{}^3R(\!h\!)\right]\nonumber \\
\Phi(k) \!\!\!&=& \!\!\!\int d^4X (\bar{E} \pi)^{\mu \nu}
n_{\mu}(X;g] \,k_{\nu}(X) \nonumber
\end{eqnarray}
where $\vec{L}^{\mu}(X;g) n_{\mu}(X;g] = 0$ and $G_{\mu \nu \rho
\sigma}$ is a covariant expression of the Dewitt metric. The
supermomentum ${\cal{H}}[ \vec{L} ]$ is smeared with a horizontal
vector field ${\cal L}$, normal to the foliation vector normal to
the leaves $n^{\mu}$; the superhamiltonian ${\cal H}\!_{\perp}[L]$
is smeared with a scalar function $L$; while the primary constraints
$p = p^i = 0$ are smeared together in a compact form, of the
constraint $\Phi(k)$, by a one-form $k_\nu$. $E$ is a kernel
function that appears first when we relate the spacetime variables
with the canonical ones; when the foliation does not depend on the
metric $E$ it equals the unit operator.


\subsection{Equivariance condition}
In order to study the explicit relation between the $\rm Diff(M)$
group and the canonical constraints, we introduce an important
mathematical restriction on the foliation, the \textsl{equivariance
condition}.

The equivariance condition follows from the requirement of general
covariance, namely that the description of the theory ought to be
invariant under changes of coordinate systems implemented by
spacetime diffeomorphisms.

A metric-dependent foliation functional
\begin{equation}
 {\cal E}:{\rm LRiem}(M)\rightarrow {\rm Fol}(M)
 \end{equation}
is defined as an \textsl{equivariant foliation} if it satisfies the
simple mathematical condition
\begin{eqnarray}
{\cal E}[f^*g]=f^{-1}\circ {\cal E}[g],
\label{Def:EquivariantFoliation2} \nonumber
\end{eqnarray}
for all Lorentzian metrics $g$ and $ f\in{\rm Diff}(M)$.

The interpretation of the condition Eq.\
(\ref{Def:EquivariantFoliation2}) is as follows: if we perform a
change of the coordinate system of the theory under a spacetime
diffeomorphism, then the expressions of the objects defined in it
will change, and so the foliation functional ${\cal E}[g]$ and the
four-metric $g$ will also change. Then, the change of the foliation
due to the change of the coordinate system must be compensated by
the change due to its functional dependence on the metric $g$. This
is essentially the \emph{passive interpretation} of the spacetime
diffeomorphisms. Loosely speaking what we have achieved with the
introduction of the equivariance condition is that the foliation
functional `looks the same' in all coordinate systems.

The physical requirement is that the change of any tensor field
$A(\cdot , g]$, associated to the foliation, under a diffeomorphism
$f$ is compensated by the change due to its functional dependence on
$g$. Hence, if we consider a diffeomorphism transformation $f$, and
we denote its pull-back operation by $f^*$, the equivariance
condition is given by the expression
\begin{eqnarray}
 (f^* A)(\cdot \,, g] = A(\cdot \,, f^* g].\nonumber
\end{eqnarray}

\subsection{Relation between the invariance groups}
One of the deepest issues to be addressed in canonical gravity is
the relation of the algebra of constraints to the spacetime
diffeomorphisms group. The canonical constraints depend on the $3+1$
decomposition and hence on the foliation functional.

The equivariance condition manifests a striking result both in its
simplicity and its implications: the action of the diffeomorphisms
group $\rm Diff(M)$ \textsl{preserves} the set of the constraints,
in the sense that it transforms a constraint into another of the
same type but of different argument. Hence, the choice of an
equivariance foliation implements that histories canonical field
variables related by diffeomorphisms are physically equivalent
\begin{eqnarray}
 \{ V_W \,, \Phi(k) \} &=&   \Phi({\cal L}_Wk)
 \label{equivVw-phi}  \nonumber\\
 \{ V_W \,, {\cal H}(\vec{L})\} &=&  {\cal H}(\delta_W \vec{L})
 \label{equivVw-Hi} \nonumber \\
 \{ V_W \,, {\cal H}_{\perp} (L)\} &=&  {\cal H}_{\perp}({\cal
 L}_WL).
 \label{equivVw-H} \nonumber
\end{eqnarray}
Here $\delta_W$ is the total change due to a diffeomorphism that
takes into account that $L^{\mu}$ is normal to $n^{\mu}$, which is
itself metric dependent.

Furthermore, this result means also that, the group $\rm Diff(M)$ is
represented in the space of the true degrees of freedom, the reduced
phase space. We can say equivalently that the space of true degrees
of freedom is invariant under $\rm Diff(M)$.

Hence, in the histories theory the requirement of the physical
equivalence of different choices of time direction is satisfied by
means of the equivariance condition.


\subsection{Reduced state space}
Finally we study the reduction procedure as implemented in the
histories framework. General relativity is a parameterized system in
the sense that it has vanishing Hamiltonian on the reduced phase
space due to the presence of first class constraints.

One may define the history constraint surface $C_h = \{ t \mapsto C,
t\in \mathbb{R}\}$ as the space of maps from the real line to the
single-time constraint surface $C$ of canonical general relativity.

The history reduced state space is obtained as the quotient of the
history constraint surface, with respect to the action of the
constraints, i.e. the space of orbits on $C_h$ arising from the
action of the constraints.

The histories Hamiltonian constraint is defined as
\begin{equation}
H_{\kappa} = \int \! dt\, \kappa (t) h_t,
\end{equation}
where $h_t := h(x_t, p_t)$ is first-class constraint. For all values
of the smearing function $\kappa (t)$, the history Hamiltonian
constraint $H_{\kappa}$ generates canonical transformations on the
history constraint surface.

It has been proved\cite{Sav03b} that the history reduced state space
$\Pi_{red}$ is a symplectic manifold that can be identified with the
space of paths on the canonical reduced state space $\Gamma_{red}$:
\begin{equation}
 \Pi_{red}= \{ t\!\mapsto \!\Gamma\!_{red},
t\!\in\!\mathbb{R}\}
\end{equation}
We have proved therefore that the histories reduced state space is
\textsl{identical} with the space of paths on the canonical reduced
state space.

Consequently the time parameter $t$ also exists on $\Pi_{red}$, and
the notion of \emph{time ordering\/} remains on the space of the
true degrees of freedom $\Pi_{red}$. This last result is in contrast
to the standard canonical theory where there exists ambiguity with
respect to the notion of time after reduction.

Moreover, the phase space action functional $S$
\begin{eqnarray} S\!:=\! \int \!\!
dt\int \!\!\! d^3 \underline{x} \! \left \{ {\tilde{\pi}}\!^{ij}
\!(t,\underline{x}) {\dot{h}}\!_{ij}\!(t,\underline{x}) \!+
\!{\tilde{p}}_i {\dot{N}}\!^i \!+ \!\tilde{p} \dot{N}\!-
\!\!{\cal{H}} \!(N)\! \!-\!\! {\cal{H}}\! (\vec{N}) \!\right\}
\nonumber
\end{eqnarray}
commutes weakly with the constraints, so it can be projected on the
histories reduced state space
\begin{eqnarray} \{ S \,, \Phi(k)+
{\cal{H}}[ \vec{L} ]+ {\cal H}\!_{\perp}[L] \}\simeq 0. \nonumber
\end{eqnarray}
It then serves its role in determining the equations of motion, as
we have shown in the theory of classical
histories\cite{observables}.

In order for a function on the full state space $ \Pi $, to be a
physical observable ({\em i.e.,\/} to be projectable into a function
on $ {\Pi}_{red}$), it is necessary and sufficient that it commutes
with the constraints on the constraint surface. Contrary to the
canonical treatments of parameterised systems, the classical
equations of motion are explicitly realised on the reduced state
space $\Pi_{red}$.

Indeed, the equations of motion are the paths on the phase space
that remain invariant under the symplectic transformations generated
by the projected action
\begin{eqnarray} \{ \tilde S
, F_t \}\,(\gamma_{cl}) = 0\,, F_t\,\, constant\,\, in \,\,t
\nonumber
\end{eqnarray}
where $\tilde{S}$ and $\tilde{V}$ are respectively the action and
Liouville functions projected on $\Pi_{red}$.

The usual dynamical equations for the canonical fields $h_{ij}$ and
$\pi^{ij}$ are equivalent to the history Poisson bracket equations
\begin{eqnarray}
\{ S \,, h_{ij} (t,\underline{x}) \}\, (\gamma_{cl}) &=& 0
            \label{eqsSh} \\[2pt]
\{ S \,, \pi^{ij} (t,\underline{x}) \}\, (\gamma_{cl}) &=& 0
\label{eqsSpi}
\end{eqnarray}
The path $\gamma_{cl}$ is a solution of the classical equations of
motion, and therefore corresponds to a spacetime metric that is a
solution of the Einstein equations.

The canonical action functional $S$ is also diffeomorphic-invariant
\begin{equation}
 \{ V_W , S \} = 0.
\end{equation}
This is a significant result: it leads to the conclusion that the
action functional and the equations of motion
(\ref{eqsSh}--\ref{eqsSpi}) are the `observables' of general
relativity theory, as has been indicated from the Lagrangian
treatment of the theory. Hence, \emph{the dynamics of the histories
theory is invariant under the group of spacetime diffeomorphisms}.

It is important to remember that the parameter with respect to which
the orbits of the constraints are defined, is not in any sense
identified with the physical time $t$. In particular, one can
distinguish the paths corresponding to the classical equations of
motion by the condition
\begin{equation}
 \{ F , \tilde{V}\}_{{\gamma}_{cl}} = 0 , \label{FV}
\end{equation}
where $F$ is a functional of the field variables, and
${{\gamma}_{cl}}$ is a solution to the equations of motion.

In standard canonical theory, the elements of the reduced state
space are all solutions to the classical equations of motion. In
histories canonical theory, however, an element of the reduced state
space is a solution to the classical equations of motion only if it
also satisfies the condition Eq.\ (\ref{FV}). The reason for this is
that the histories reduced state space ${\Pi}_{red}$ contains a much
larger number of paths (essentially all paths on ${\Gamma}_{red}$ ).
For this reason, histories theory may naturally describe observables
that commute with the constraints but which are not solutions to the
classical equations of motion.

This last point should be particularly emphasised, because of its
possible corresponding quantum analogue. We know that in quantum
theory, paths may be realised that {\em are not\/} solutions to the
equations of motion. My belief is that the histories formalism will
distinguish between instantaneous laws\cite{egregium} (namely
constraints), and dynamical laws (equations of motion). Hence, it is
possible to have a quantum theory for which the instantaneous laws
are satisfied, while the classical dynamical laws are not. This
distinction is present, for example, in the history theory of the
quantised electromagnetic field, where all physical states satisfy
the Gauss law exactly, however electromagnetism field histories are
possible which do not satisfy the dynamical equations, i.e.,
Maxwell's equations. For parameterised systems, this distinction is
not possible within the canonical formalism, nevertheless as we
explained, it does arise in the histories formalism.

The equations of motion (\ref{FV}) imply that physical observables
have constant values on the solutions to the classical equations of
motion. This need not be the case quantum mechanically, hence
quantum realised paths need not be characterised by `frozen' values
of their physical parameters.


\section{Notes on quantization}

These are significant results for developing a theory of quantum
gravity. It indicates that the histories scheme can incorporate
intrinsically the basic symmetry of general relativity, namely
general covariance---as manifested by the existence of a realisation
of the $\rm Diff(M)$ group---and the invariance of the Hamiltonian
constraints under its action.

Furthermore it provides a possible quantum gravity theory where the
full Lorentzian metric may be quantised, unlike some spatial part of
the metric of the canonical schemes. For this purpose we may follow
the quantum algorithm we described in the beginning of this
presentation. That is to seek a representation of the history
algebra or the histories commutation relations that are defined with
reference to the whole of spacetime and not just a 3-surface; in
particular, these history variables include a quantised Lorentzian
spacetime metric. Of course problems of defining properly quantum
Hamiltonian operators still remains in a first estimation of the
formalism.

Another possible direction to follow is to develop a histories
analogue of loop quantum gravity, as this is a successful canonical
theory in many respects. In the canonical treatment, the basic
algebra is defined with reference to objects that have support on
loops in the three-dimensional surface $\Sigma$.  The natural object
in the histories description is the $SL(2,{\bf C})$) connection. An
obvious first step would be to write the kinematical Hilbert space
based on a representation of an $SL(2,C)$ connection on $M$ instead
of the $SU(2)$ connection on $\Sigma$.

This is a more complicated matter, and there is no guarantee that
there exists a correspondence between the histories $SL(2,{\bf C})$
theory and the canonical $SU(2)$ one. The major difference is that
the $SL(2,{\bf C})$ group is non-compact, hence the definition of
the proper Hilbert space cannot follow the steps of the canonical
construction.

 The mathematical structures of a quantisation based on
histories will conceivably be very different from those in the
canonical theory. For this reason, the history construction may
uncover substantially different properties from those that arise in
the existing approaches to loop quantum gravity.

\section{Acknowledgments}
This work was supported by a Marie Curie
 Reintegration Grant of the European Commission.
\bibliography{apssamp}


\end{document}